\documentclass[twocolumn,aps,prb,showpacs,tightenlines,amsmath,amssymb,superscriptaddress]{revtex4}
\usepackage{graphicx}    
\usepackage{amssymb}  
\usepackage{dcolumn}
\usepackage{amsmath}     
\usepackage{bm}                                          
\usepackage{colordvi}                   

\begin{document}             
\title{Electron spin relaxation in GaAs$_{1-x}$Bi$_x$: Effects of spin-orbit 
tuning by Bi incorporation}
\author{H. Tong}
\affiliation{Hefei National Laboratory for Physical Sciences at Microscale and
  Department of Physics, University of Science and Technology of China, Hefei,
  Anhui, 230026, China} 
\author{X. Marie}
\affiliation{Universit\'{e} de Toulouse, INSA-CNRS-UPS, LPCNO, 135 avenue 
de Rangueil, F-31077 Toulouse, France} 
\author{M. W. Wu}
\thanks{Author to whom correspondence should be addressed}
\email{mwwu@ustc.edu.cn.}
\affiliation{Hefei National Laboratory for Physical Sciences at Microscale and
  Department of Physics, University of Science and Technology of China, Hefei,
  Anhui, 230026, China} 
\date{\today}

\begin{abstract}
The electron spin relaxation in $n$-type and intrinsic GaAs$_{1-x}$Bi$_x$ with
Bi composition $0\le x \le 0.1$ is investigated from the microscopic kinetic
spin Bloch equation approach. The incorporation of Bi is shown to markedly
decrease the spin relaxation time as a consequence of the modification of the
spin-orbit interaction. We demonstrate that
the density and temperature dependences of spin relaxation time in
GaAs$_{1-x}$Bi$_x$ resemble the ones in GaAs. 
Meanwhile, the Bir-Aronov-Pikus mechanism is
found to be negligible compared to the D'yakonov-Perel' mechanism in 
intrinsic sample. Due to the absence of direct
measurement of the electron effective mass in the 
whole compositional range under investigation, 
we further explore the effect of a possible variation of electron effective mass
on the electron spin relaxation. 
\end{abstract}



\maketitle

Recently, the highly mismatched semiconductor alloys have received considerable
interest due to their unusual compositional dependences of electronic and 
spintronic
properties.\cite{Fran1,Fran2,Fran3,Fran4,Alberi1,Alberi2,Hwang,Deng,Usman} 
These alloys are typically obtained by 
isoelectronic doping of very light or very heavy
elements into the traditional 
binary. For example, dilute incorporation of N or Bi into GaAs is found to 
induce a strong reduction of the band
gap.\cite{Fran1,Fran2,Fran3,Alberi1,Alberi2}  
This property makes them a focus of current researches in long-wavelength
optoelectronics and shows great potential to work as highly efficient solar cell
and thermoelectric power generators.\cite{Bertulis,Dimroth,Hwang,Fran2,Alberi1,Alberi2,Fran3}
The alloys GaAs$_{1-x}$Bi$_x$, which have been less investigated than
GaAs$_{1-x}$N$_x$, have great potential for device
applications.\cite{Hwang,Cooke,Fluegel} One reason  
is that the incorporation of Bi 
modulates the band gap primarily through distorting the valence-band edge and
leaves the conduction band almost unchanged, while N in GaAs mainly modulates
the conduction-band edge and results in a substantial degradation of electron 
mobility.\cite{Zhang,Cooke,Fluegel,Deng} Moreover, a strong 
enhancement of spin-orbit splitting energy is 
observed in GaAs$_{1-x}$Bi$_x$.\cite{Fran4,Alberi1,Alberi2} This indicates a
new way of tailoring the spin-orbit coupling (SOC) 
through Bi incorporation and hence suggests its application 
in the field of spintronics.\cite{Fran4,Alberi1,Alberi2} 
A systematical calculation of the electron spin lifetime in GaAs$_{1-x}$Bi$_x$
is of fundamental importance  for   
both theoretical and application purposes.  

In this work, we employ the microscopic kinetic spin Bloch equation (KSBE)
approach,\cite{Wu1,Wu2,Wu3,Wu5} which has been 
demonstrated to be successful in the study of spin dynamics in
semiconductors,\cite{Wu5} to calculate the electron spin relaxation time
(SRT) in $n$-type and intrinsic GaAs$_{1-x}$Bi$_x$. We focus on the electron
spin relaxation from the D'yakonov-Perel' (DP) mechanism\cite{DP} in $n$-type
case and further take into account the Bir-Aronov-Pikus (BAP)
mechanism\cite{BAP} in the intrinsic condition. Both mechanisms are shown to be
pronouncedly enhanced in GaAs$_{1-x}$Bi$_x$ and result in enhanced spin
relaxation rate when Bi composition increases. Meanwhile, the BAP mechanism is
recognized to be still negligible in the intrinsic 
case as in GaAs.\cite{Jiang} The density and temperature dependences of the SRTs are
explicitly studied, where the features are found to resemble those in GaAs.\cite{Jiang} In
addition, we discuss the impacts of possible variations of the electron and/or  
hole effective masses on the spin relaxation in GaAs$_{1-x}$Bi$_x$.

The KSBEs derived from the nonequilibrium Green
function method reads\cite{Wu1,Wu2,Wu5}
\begin{equation}
 \left. \left. \partial_t \rho_{\bf k}=
\partial_t\rho_{\bf k}\right|_{\rm coh}+\partial_t\rho_{\bf k}\right|_{{\rm scat}}, \label{KSBE1}
\end{equation}
in which $\rho_{\bf k}$ is the single-particle density matrix of electrons with its
diagonal term representing the distribution of each spin band and the
off-diagonal term denoting the correlation between the two spin bands. The
coherent terms $\partial_t\rho_{\bf k}|_{\rm coh}$ describe the spin
precession of electrons due to the Dresselhaus SOC\cite{Dresselhaus} as well as
the Hatree-Fock Coulomb interaction. $\partial_t\rho_{\bf k}|_{{\rm scat}}$
stands for the scattering terms, including the electron-impurity,
electron-phonon, electron-electron and electron-hole Coulomb scatterings, and
also the electron-hole exchange scattering. The explicit expressions of both
coherent and scattering terms can be found in Ref.~\onlinecite{Jiang}.

\begin{figure}[htb]
  \begin{center}
    \includegraphics[width=8cm]{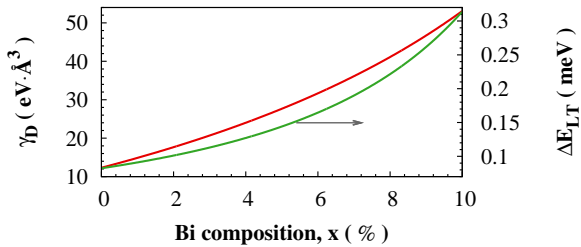}
  \end{center}
  \caption{ The SOC parameter
    $\gamma_D$ and the longitudinal-transverse splitting $\Delta E_{\rm LT}$ as
    function of Bi composition. Note that the scale of $\Delta E_{\rm LT}$ is on
  the right hand side of the frame.} 
  \label{figtw1}
\end{figure}

The incorporation of Bi influences the spin relaxation via both the DP and BAP
mechanisms through the modulations of the band gap $E_g$ and the spin-orbit
splitting $\Delta_{\rm SO}$. The compositional dependences of $E_g$ and
$\Delta_{\rm SO}$ are obtained by linear fittings of the experimental results
summarized in Fig.~3(c) of Ref.~\onlinecite{Alberi2}: $E_g=(1.435-7.0x)$~eV, and
$\Delta_{\rm SO}=(0.329+5.27x)$~eV. The electron effective mass $m_e^\ast$ is
taken to be fixed to its value in GaAs 
$m_{\rm GaAs}$, which is supported by the direct Shubnikov-de Haas measurement
of $m_e^\ast$\cite{Fluegel} and also the mobility experiment.\cite{Cooke} 
The coefficient of the Dresselhaus SOC takes
the form $\gamma_D=(4/3)(m^\ast_e/m_{\rm
  cv})(1/\sqrt{2{m_e^\ast}^3E_g})(\eta/\sqrt{1-\eta/3})$, in which
$\eta=\Delta_{\rm SO}/(E_g+\Delta_{\rm SO})$, and $m_{\rm cv}$ is a constant 
close in magnitude to the free-electron mass $m_0$.\cite{Aronov} Besides, the
longitudinal-transverse splitting, which is a scale of the  
electron-hole exchange interaction (we neglect the short-range electron-hole
exchange interaction due to its marginal effect\cite{Jiang}), is given  as
$\Delta E_{\rm LT}={2e^2\hbar^2E_P}/({3\pi\epsilon_0\kappa m_0a_{\rm
    Bohr}^3E_g^2})$. Here $a_{\rm Bohr}$ denotes the exciton Bohr radius in
bulk, $\kappa$ is the static dielectric constant, and $E_P$ is the band
parameter.\cite{Tong} In GaAs$_{1-x}$Bi$_x$, $\kappa$, $E_P$ and also the hole
effective mass (which influences $a_{\rm Bohr}$ together with the electron-hole
Coulomb scattering) are still not well investigated and are assumed to be the
values in GaAs.   In Fig.~\ref{figtw1}, we plot the Bi-compositional dependences
of $\gamma_D$ and 
$\Delta E_{\rm LT}$. It is seen that both $\gamma_D$ and 
$\Delta E_{\rm LT}$ are markedly enhanced with the increase of Bi composition and
reach up to roughly four times of their original values when $x=10$~\%. Therefore,  
the electron spin relaxation from both the DP and BAP mechanisms is expected to be 
remarkably strengthened thanks to the incorporation of Bi in GaAs. However,
considering that the BAP mechanism is demonstrated to be negligible compared
with the DP mechanism in the intrinsic GaAs,\cite{Jiang} its effect is still 
expected to be marginal in the presence of Bi incorporation. In the following, we
calculate the SRT in $n$-type and 
intrinsic GaAs$_{1-x}$Bi$_x$ from the KSBE approach with all scatterings
explicitly included. 

\begin{figure}[htb]
  \begin{center}
    \includegraphics[width=7.5cm]{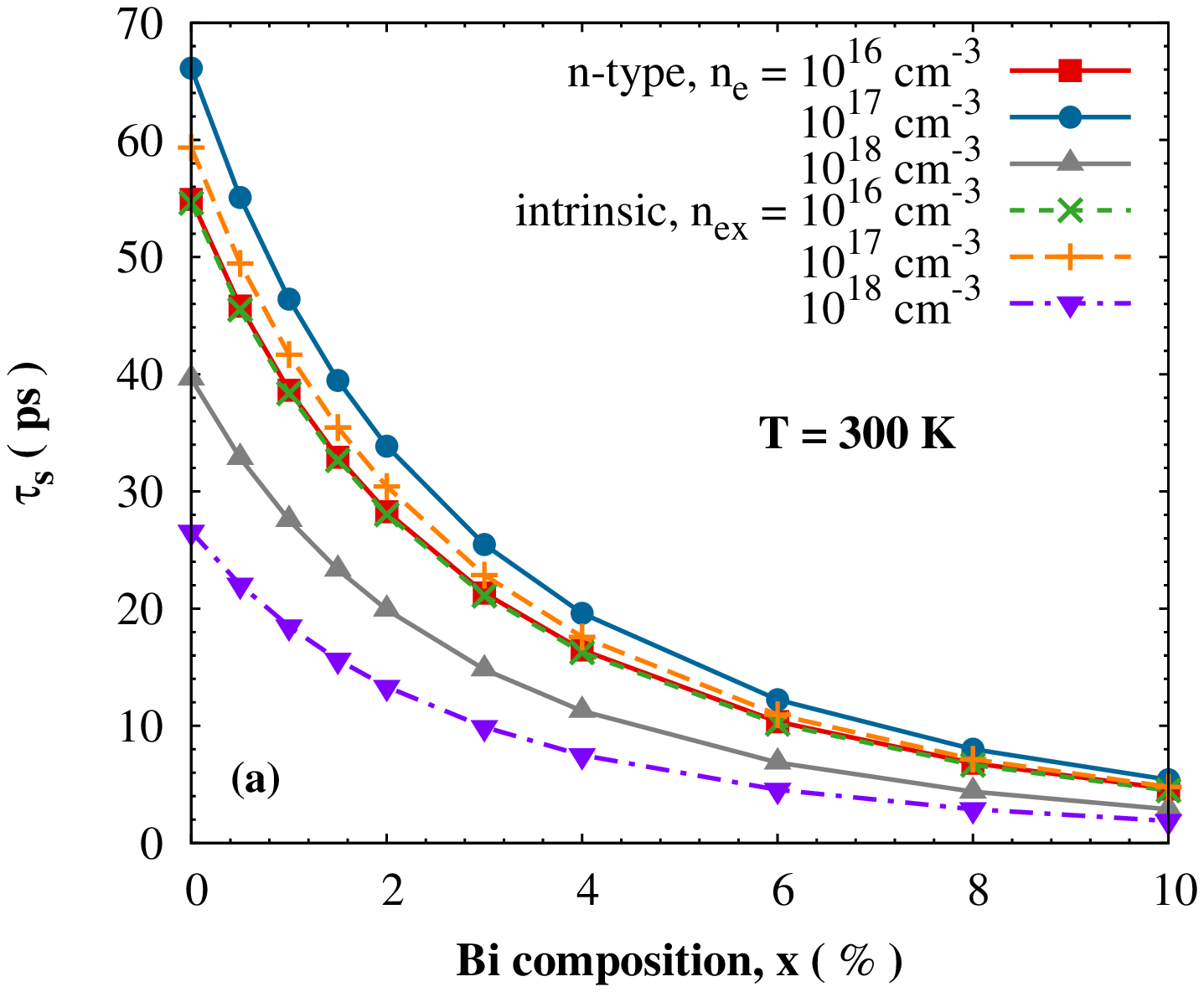}
    \includegraphics[width=7.5cm]{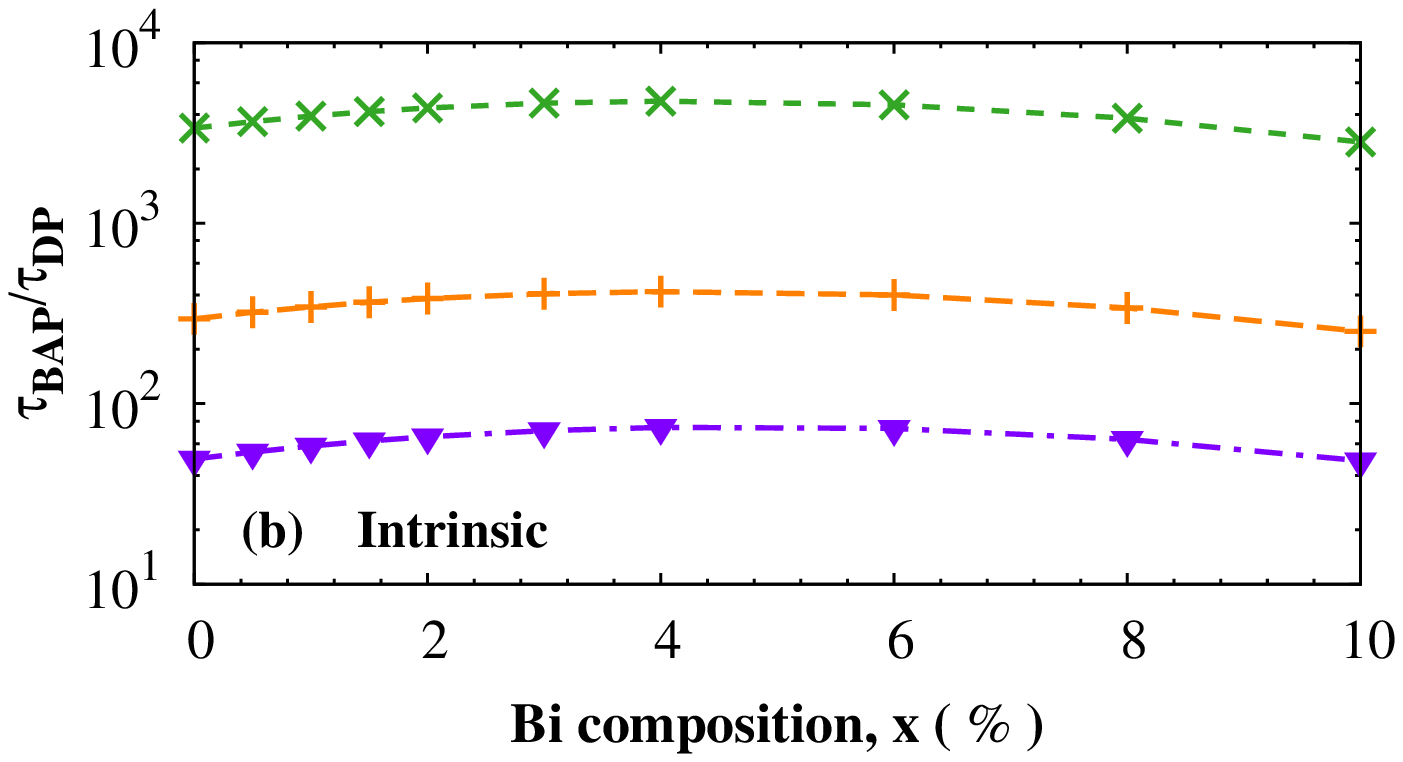}
  \end{center}
  \caption{ (a) SRT $\tau_s$ as function of Bi composition at temperature
    $T=300$~K. For $n$-type condition with electron densities
    $n_e=10^{16}$, $10^{17}$ and $10^{18}$~cm$^{-3}$ (solid curves) and for
    intrinsic condition with photoexcitation densities $n_{\rm ex}=10^{16}$,
    $10^{17}$ and $10^{18}$~cm$^{-3}$ (dashed curves). (b) The ratio of the SRT
    due to the BAP mechanism $\tau_{\rm BAP}$ to that due to the DP mechanism
    $\tau_{\rm DP}$ corresponding 
to the intrinsic condition in (a). Note that the same
  color and type of line and point are used for the corresponding
   photoexcitation density. } 
  \label{figtw2}
\end{figure}

We first investigate the Bi-compositional dependence of the SRTs in
GaAs$_{1-x}$Bi$_x$. In Fig.~\ref{figtw2}, the SRTs as function of the Bi
composition are plotted for both $n$-type and intrinsic conditions at
$T=300$~K. We consider three electron densities from $n$-doping or
photoexcitation, i.e., $10^{16}$, $10^{17}$ and $10^{18}$~cm$^{-3}$,
corresponding to the nondegenerate, intermediate, and degenerate regimes,
respectively. One observes that the SRTs significantly decrease 
with  the increase  of Bi incorporation. In all conditions, the SRTs
decrease with the increase of Bi composition and reach  values around one
order of magnitude smaller than the corresponding 
Bi-free ones at $x=10$~\%. 
Another feature shown in Fig.~\ref{figtw2} is that, for both $n$-type (solid
curves) and intrinsic (dashed curves) cases, the SRTs first increase 
and then decrease with increasing electron
densities. This nonmonotonic density dependence of the SRT resembles the one in
GaAs and is attributed to the crossover from the nondegenerate to 
degenerate limit.\cite{Jiang,Tong2} 
It is also noticed that the curves corresponding to the $n$-type and
  intrinsic cases at electron density $10^{16}$~cm$^{-3}$ coincide with each
other. The underlying physics is that at high temperature and low electron
density, the electron-phonon scattering is dominant while all other scatterings
are irrelevant. With the increase of the electron density and hence also the
corresponding impurity or hole density, the electron-impurity and electron-hole
scatterings are strengthened, which leads to a considerable difference between the
$n$-type and intrinsic cases. Accordingly, the differences of the corresponding
curves with the same electron density are observed.
We further address the relative importance of the DP and BAP mechanisms in intrinsic
GaAs$_{1-x}$Bi$_x$. 
In Fig.~\ref{figtw2}(b), we plot the ratio of the SRT due to
the BAP mechanism, $\tau_{\rm BAP}$, to that due to the DP mechanism, $\tau_{\rm
  DP}$, in the conditions corresponding to  Fig.~\ref{figtw2}(a). It is seen that
$\tau_{\rm BAP}/\tau_{\rm DP}$ is insensitive to the composition of
Bi, which results from the comparable increasing rates of $\gamma_D$ and $\Delta
E_{\rm LT}$ with Bi-doping as shown in Fig.~\ref{figtw1}. Meanwhile, $\tau_{\rm BAP}/\tau_{\rm DP}$
increases with increasing electron density but remains larger than
$40$. Thus the BAP mechanism is negligible in intrinsic
GaAs$_{1-x}$Bi$_x$ as expected.\cite{Jiang}
Furthermore, taking into consideration  the experimental evidence of an 
increase of the hole effective mass in 
GaAs$_{1-x}$Bi$_x$,\cite{Alberi2,Tiedje,Narg} 
we perform the calculation of the SRT in intrinsic GaAs$_{1-x}$Bi$_x$ by
doubling the heavy- and light-hole effective masses. The obtained SRTs for all three
photoexcitation densities remain almost unchanged, thanks to the marginal effects of 
the electron-hole Coulomb and exchange scatterings at room temperature.

\begin{figure}[thb]
\begin{minipage}[]{10cm}
\hspace{-1.8cm}
      \includegraphics[width=4.5cm]{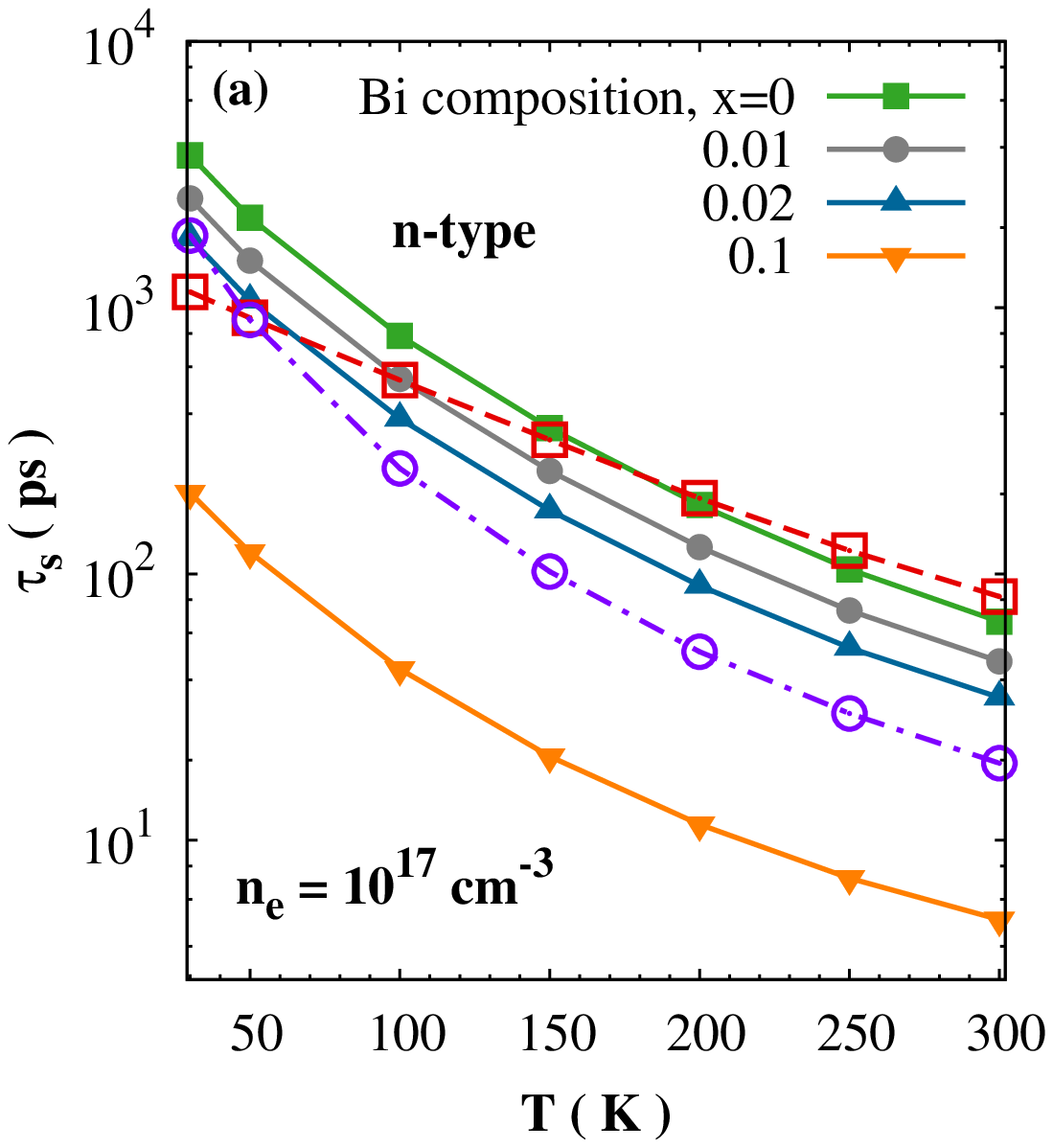}
\hspace{-0.5cm}
      \includegraphics[width=4.5cm]{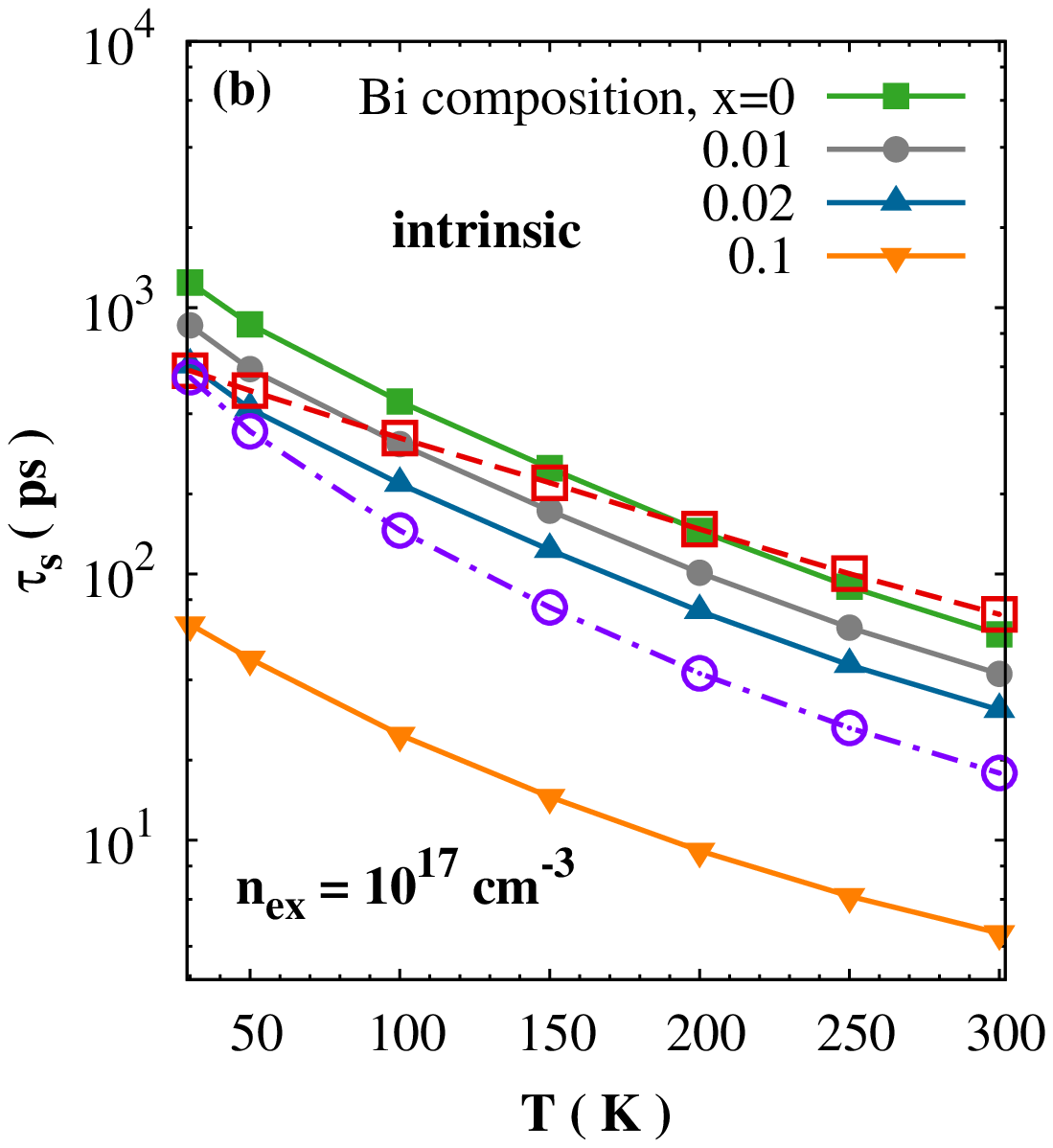}
  \end{minipage}
  \begin{center}
    \includegraphics[width=7cm]{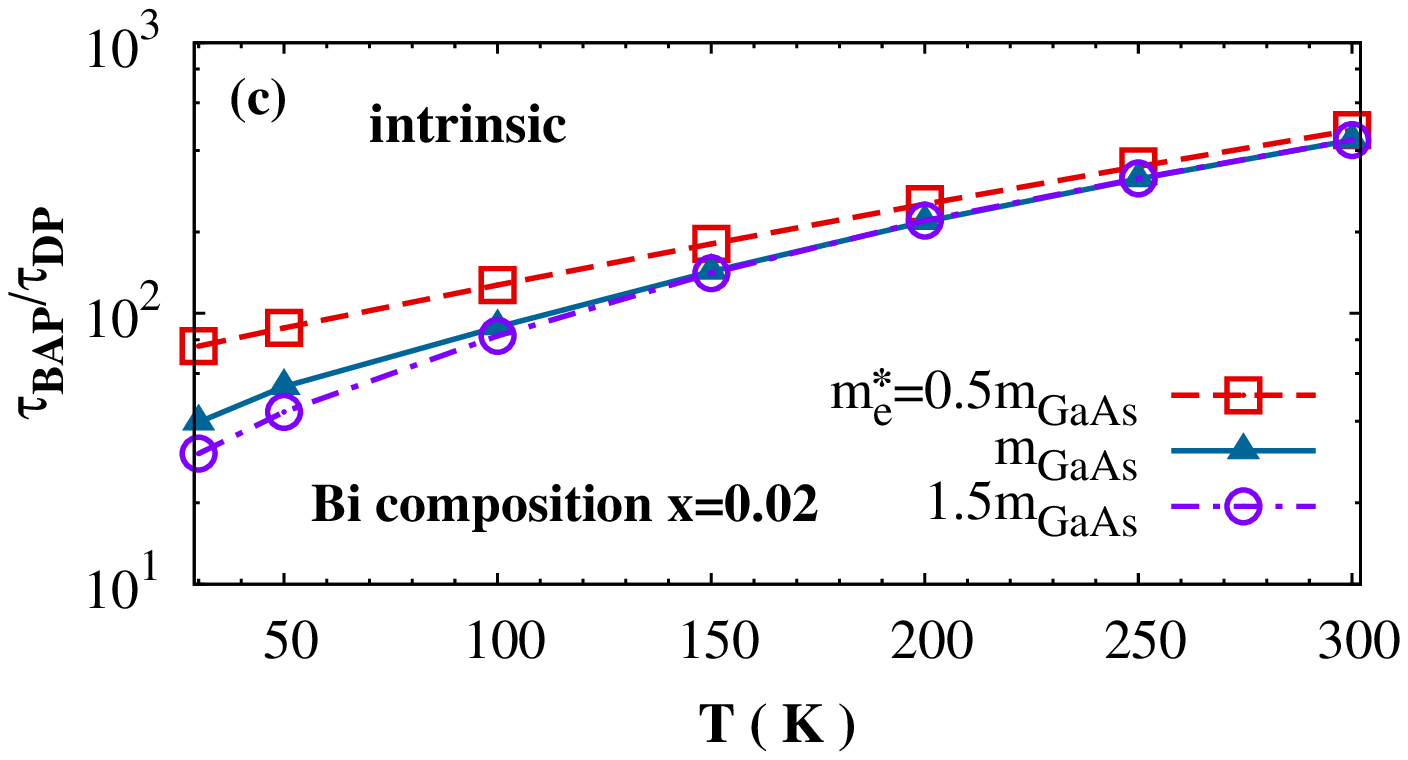}
  \end{center}
  \caption{ SRT $\tau_s$ as function of temperature for four  Bi compositions:
  $x=0$, $0.01$, $0.02$ and $0.1$. (a) $n$-type condition with electron density
  $n_e=10^{17}$~cm$^{-3}$ and (b) intrinsic condition with photoexcitation
  density $n_{\rm ex}=10^{17}$~cm$^{-3}$. The STRs in GaAs$_{0.98}$Bi$_{0.02}$
  calculated by artificially using 
electron effective masses $0.5m_{\rm GaAs}$ (dashed curves with $\square$) and
  $1.5m_{\rm GaAs}$ (dashed curves with $\odot$) are also plotted.
(c) The ratio of the SRT due to the BAP 
mechanism $\tau_{\rm BAP}$ to that due to the DP mechanism
   $\tau_{\rm DP}$ in intrinsic GaAs$_{0.98}$Bi$_{0.02}$ for three electron effective masses. Note that the same
  color and type of line and point are used for the corresponding $m_e^\ast$
 in (b).} 
  \label{figtw3}
\end{figure}

We then investigate the temperature dependences of the SRTs in both $n$-type
and intrinsic GaAs$_{1-x}$Bi$_x$ with the electron density $10^{17}$~cm$^{-3}$
for different Bi compositions $x=0$, $0.01$, $0.02$ and $0.1$. One observes from
Figs.~\ref{figtw3}(a) and \ref{figtw3}(b) that in 
all cases, the SRTs decrease monotonically with the increase of temperature. This
behavior has formerly been studied both experimentally\cite{Kikk,Oertel} and
theoretically\cite{Jiang,Tong2} in GaAs and the underlying 
physics lies in the drastic
increase of the inhomogeneous broadening\cite{Wu1,Wu2} due to the ascending
temperature. Moreover, the giant tailoring of SRTs resulting from the
incorporation of Bi at all temperatures is seen in the figures. 

\begin{table}[thb]
\caption{Coefficient of the Dresselhaus SOC $\gamma_D$ and 
  longitudinal-transverse exchange splitting $\Delta E_{\rm LT}$ 
in GaAs$_{0.98}$Bi$_{0.02}$ for different electron effective masses. }
\begin{tabular}{llllllllll}\hline\hline
  $m_e^\ast$ ($m_{\rm GaAs}$)&\ \ \ &  $0.5$  &\ \ \ \ \ \ &  1.0 &\ \ \ &  1.5  &&  \\ 
  $\gamma_D$ (eV$\cdot$ \r{A}$^3$) &      &  24.98        &   &  17.67    &      & 14.43&&      \\
  $\Delta E_{\rm LT}$ (meV) &      &  0.02    &           & 0.10     &      &0.22 &&      \\
\hline\hline
\end{tabular}\\
\end{table}

It is noted that the electron effective mass in GaAs$_{1-x}$Bi$_x$ from a direct 
measurement is only available for Bi composition up to $x=0.01$.\cite{Fluegel}
Although it is argued to stay untouched under Bi incorporation,\cite{Fluegel,Cooke,Hwang}
there are also works suggesting an increasing\cite{Pett,Mbarki} or
a nonmonotonic\cite{Pettinari} behavior with the increase of Bi
composition. Therefore, it is meaningful to perform an investigation to include the
possible modification of the electron effective mass  in GaAs$_{1-x}$Bi$_x$. For
simplicity, we calculate  
the temperature dependence of the SRTs in both $n$-type and intrinsic
GaAs$_{0.98}$Bi$_{0.02}$ by artificially setting the electron effective
mass $m_e^\ast=0.5m_{\rm GaAs}$ and $1.5m_{\rm GaAs}$. The possible variation of
the electron effective mass is reasonably covered within this
range.\cite{Pettinari,Mbarki} We indicate the obtained SRTs with dashed curves
in Figs.~\ref{figtw3}(a) and \ref{figtw3}(b), accordingly. It is seen that, the
variation of $m_e^\ast$ markedly changes the specific values, together with the
temperature dependence of the SRTs. In order to elucidate this behavior, we
calculate the variation of $\gamma_D$ and $\Delta E_{\rm LT}$ from the change of
electron effective mass.  From Table.~I, one notices that $\gamma_D$ visibly
decreases with the increase of $m_e^\ast$, while the reverse is true for $\Delta  
E_{\rm LT}$. One may hence expect the BAP mechanism to play a role in the spin
relaxation in intrinsic GaAs$_{1-x}$Bi$_x$ if $m_e^\ast$ does increase. However, 
a detailed calculation of $\tau_{\rm BAP}/\tau_{\rm DP}$ shown in
Fig.~\ref{figtw3}(c) indicates that the BAP mechanism is still negligible in the
intrinsic case. This is analyzed as 
follows. The overall coefficient of the long-range electron-hole exchange
scattering is actually proportional to $a_{\rm Bohr}^3\Delta E_{\rm LT}$ rather
than $\Delta E_{\rm LT}$ alone,\cite{Jiang} and hence is independent of
$m_e^\ast$. Therefore, only slight change of $\tau_{\rm BAP}/\tau_{\rm
  DP}$ with the increasing $m_e^\ast$ is observed in Fig.~\ref{figtw3}(c).
We then focus on the DP mechanism. The variation of the
 electron effective mass influences the DP
spin relaxation in  two ways: (i) it modulates the coefficient of 
the Dresselhaus spin-orbit
coupling $\gamma_D$; (ii) it changes the energy spectrum together with the
electron distribution and hence modifies both the inhomogeneous broadening and
the scattering. With the increase of $m_e^\ast$, on one hand, $\gamma_D$
decreases and hence tends to suppress the spin relaxation; on the other hand,
the electrons are more easily thermalized to high $k$ states in the case of a
flatter energy spectrum. This results in a larger inhomogeneous broadening and hence
tends to enhance the spin relaxation (which is obvious when the change of the
inhomogeneous broadening overcomes that of the scatterings). At high
temperature, the latter factor dominates, so larger 
$m_e^\ast$ implies smaller $\tau_s$. At low temperature, the situation is
reversed. Accordingly, the crossovers are seen in Figs.~\ref{figtw3}(a) and 
\ref{figtw3}(b) [although less obvious in Fig.~\ref{figtw3}(b)]. 

In summary, we have investigated the electron spin relaxation in $n$-type
and intrinsic GaAs$_{1-x}$Bi$_x$ with Bi composition $0\le x \le 0.1$ by
applying the fully microscopic KSBE approach. The variations of the band gap and
the spin-orbit splitting energy are explicitly considered. We show that the
electron spin relaxation time decreases significantly by the incorporation of
Bi, which suggests potential 
applications of GaAs$_{1-x}$Bi$_x$ in the engineering of spintronic devices. The
density and temperature dependences of the SRT are studied for different Bi
compositions, with their behaviors found to resemble the ones in
GaAs. Meanwhile, we show that the BAP mechanism is
negligible compared to the DP mechanism in the
intrinsic condition. 
Moreover, due to the absence of unambiguous 
experimental data of $m_e^\ast$ in the whole
compositional range under investigation and also the existing contradictions in
the literature, we perform calculations to include the possible variation of
$m_e^\ast$ with the change of Bi incorporation. Profound impacts on the explicit value
together with the temperature  
dependence of the SRT are found. This makes the precise measurement of
$m_e^\ast$ in GaAs$_{1-x}$Bi$_x$ highly desirable from both theoretical and
engineering points of view.

This work was supported by the National Basic Research 
Program of China under Grant No. 2012CB922002 and 
the National Natural Science Foundation of China under Grant No. 10725417.

\end{document}